\documentclass[aps,prb,twocolumn,epsf,epsfig,amsmath]{revtex4}
\usepackage{latexsym}
\usepackage{hyperref}
\usepackage{times}
\usepackage{graphicx}
\usepackage{amsmath}
\usepackage{dcolumn}
\usepackage{bm}
\usepackage{textcomp}
\usepackage{color}
\usepackage{amssymb}
\usepackage{xcolor}
\usepackage{float}
\usepackage{array}
\usepackage[export]{adjustbox}

\begin{document}
\title{Identification of dark axisymmetric plasma modes in partially gated two-dimensional electron gas disk}
\author{M.V. Cheremisin}
\affiliation{A.F.Ioffe Physical-Technical Institute, St.Petersburg, Russia}
\date{\today}
\begin{abstract}
We investigate the dark axisymmetric plasmon spectrum for partially gated two-dimensional gas of the disk shape. The extension of the central gate spot provides a change in plasmon dispersion from the root to a linear dependence on the wave vector. Intriguingly, the interaction between neighboring modes occurs because of stepwise change in carrier screening across the circumference of the gate. This behavior is unexpected when changing from a fully screened to unscreened two-dimensional gas by increasing the dielectric width under the gate [A.L.Fetter, Phys.Rev.B 33, 5221 (1986)]. Our results provide the accurate identification of axisymmetric plasmon modes recently observed in experiment and, in addition, pave the way for feedback plasmon resonator study.
\end{abstract}
\maketitle

Plasma oscillations in two-dimensional(2D) electron gas were first predicted in the 60s by F.Stern\cite{Stern67} for ungated and, then analyzed for
gated systems\cite{Chaplik72}. Over the next decade the plasmon assisted infrared absorbtion\cite{Grimes76,Allen77,Theis78,Heitmann82}
and emission\cite{Tsui80} has been reported. Since the Stern's pioneering discovery the enormous efforts were done to clarify the plasmon behavior found to be influenced by magnetic field\cite{Chiu74,Nakayama74,Volkov88}, retardation effects\cite{Kukushkin03}, sample geometry\cite{Fetter86,Kosevich88,Zabolotnykh22} and quality\cite{Falko89,Cheremisin17}.

A decade ago it was shown\cite{Burke00,Rana08,Aizin12} that the plasma oscillations in two-dimensional systems with arbitrary attached gates and contacts can be elegantly described in terms of classical theory of electrical circuits with distributed parameters. The proposed quasi-static LC approach is extremely powerful for description of plasmon excitations in quasi one-dimensional stripes of 2D gas with periodic grating\cite{Aizin13}. Highly motivated by recent experiments dealt with axisymmetric plasmon excitations observed\cite{Kukushkin17,Kukushkin18,Kukushkin23} in partially gated 2D disk we attempt to probe LC approach in this case. Indeed, the axisymmetric modes known\cite{Fetter86,Zagorodnev21,Rodionov20} to have zero angular momentum $l=0$ and, therefore are called "dark" because of weak coupling with external electromagnetic fields. Arguing that axisymmetric modes are purely radial we will use one-dimensional LC-approach\cite{Aizin12} in order to find the plasmon spectra and, then compare theory predictions with recent experimental data. In present study we will disregard the anomalous Goos-H\"{a}nchen phase shift\cite{Nikitin14,Kang17} of incident and reflected plasma waves at the sharp edge of 2D gas. Then we will confirm the validity of the current use of the quasi-static LCD approach for the analysis of real experimental data.

\begin{figure}[tbp]
\begin{center}\leavevmode
\includegraphics[width=1.0\linewidth]{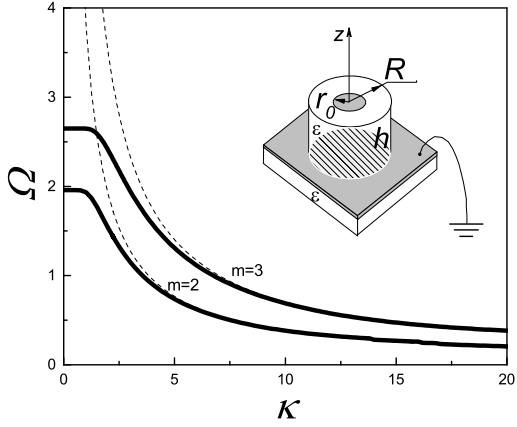} \caption[]{\label{Fig1} Inset: the experimental setup. Main panel: the dimensionless frequency $\Omega=\omega/\omega_{0}(\varepsilon)$ vs screening parameter $\kappa=\sqrt{R/2h}$ for lowest plasma modes with $\alpha_{02}=3.831$ and $\alpha_{03}=7.016$ respectively for totally covered atop 2DEG, i.e. when $r_{0}=R$. Dashed curves represent the asymptotes $\Omega=\frac{\alpha_{0m}}{\kappa}$ for highly screened 2D gas.}
\end{center}
\end{figure}

Let us consider two-dimensional electron system of disk geometry depicted by shaded area in Fig.\ref{Fig1}, inset. The 2D gas mesa of radius $R$ is grounded peripherally and, moreover, embedded into an homogeneous environment of a dielectric constant $\varepsilon$. The cylindrical slab of the height $h$ is covered atop by the gate of radius $r_{0} \leq R$. The device is surrounded by air. For clarity, we will assume that the radius $R$ of 2DEG is fixed. In contrast, the slab height $h$ and the gate radius $r_{0}$ may vary. Let us first consider the simple case of totally gated 2D gas, i.e when $r_{0}=R$ and variable height $0 < h < \infty$ of the slab.

In general, the LC approach\cite{Aizin12} is based on hydrodynamic model\cite{Fetter86} which includes the Euler equation and the continuity equation for two-dimensional electron gas. Instead of a complete set of Maxwell's equations, taking into account the retardation\cite{Rodionov20}, a quasi-static Poisson equation is used to find the in-plane potential of a two-dimensional gas. In what follows we will demonstrate that retardation effects are insignificant for the actual experimental data\cite{Kukushkin17,Kukushkin18,Kukushkin23}. According to Ref.\cite{Fetter86} the set of equations constituent the hydrodynamic model could be linearized with respect to small amplitude of plasma wave excitations. Fortunately, the model can be greatly simplified down to so-called telegrapher's equations \cite{Aizin12} for radial components of in-plane potential $\tilde{U}$ and the current $\tilde{I}$:
\begin{eqnarray}
\frac{\partial \tilde{U}}{\partial r}=-L \frac{\partial \tilde{I}}{\partial t},\qquad \qquad \qquad
\label{Telegraph_1}\\
\frac{\partial \tilde{I}}{\partial r}=-C \frac{\partial \tilde{U}}{\partial t}, \qquad \qquad \qquad
\label{Telegraph_2}\\
L(r)=\frac{m^{*}}{e^{2}n} \frac{1}{2\pi r}, \qquad C(r)=\frac{(1+\coth(qh))\varepsilon qr}{2}.
\label{L_C}
\end{eqnarray}
Here, $L,C$ are the inductance and capacitance per unit length respectively, $m^{*}$ and $n$ are the effective mass and density of 2D carriers. Then, $q$ is the plasmon wave vector originating from solution of the Poisson equation taking into account the actual dielectric environment.

We now search the solution of Eq.(\ref{Telegraph_1},\ref{Telegraph_2}) separating the temporal and spatial components for potential $\tilde{U}=U(r)\exp(i\omega t)$ and
current $\tilde{I}=I(r)\exp(i\omega t)$. Finally, the equation for radial part of the potential $U(r)$ yields:
\begin{equation}
\frac{d^{2}U}{d\rho^{2}}+\frac{1}{\rho}\frac{dU}{d\rho}+U=0.
\label{Bessel}\\
\end{equation}
Here, we introduced the dimensionless radius $\rho=r/\lambda$, where $\lambda=\frac{1}{\omega\sqrt{LC}}$ is the length scale of the problem.
The general solution of Eq.(\ref{Bessel}) is given by the sum of zero order Bessel functions of first(second) kind, namely
\begin{equation}
U(\rho)=AJ_{0}(\rho)+BY_{0}(\rho).
\label{Solution_Bessel}\\
\end{equation}
The arbitrary constants in Eq.(\ref{Solution_Bessel}) can be found by use of a certain boundary conditions. Arguing the potential is finite in the disk center $\rho \rightarrow 0$ we put $B=0$ to avoid divergent behavior of the second term in Eq.(\ref{Solution_Bessel}). The condition of zero current at the disk center $\rho=0$ is fulfilled identically since $I \sim \rho \frac{\partial U}{\partial \rho}\mid_{0}\equiv 0$. The outer boundary condition on the circumference of the disk requires a detailed analysis. Recent studies\cite{Nikitin14,Kang17} clearly demonstrate an anomalous Goos-H\"{a}nchen phase shift of incident and reflected plasma waves caused by presence of near-field evanescent waves at the sharp edge of 2D crystal. Although this effect is indeed fundamental, the present scheme of a grounded 2D system\cite{Kukushkin17,Kukushkin18,Kukushkin23} relevant in our case does not meet the criterion of a sharp edge of 2D crystal. Therefore, we will assume a simple case of zero peripheral current $I(R)\sim J_{0}'(R/\lambda)=0$. The later gives the dispersion equation for plasmon excitations
\begin{equation}
\omega \sqrt{LC} R=\alpha_{0m}.
\label{Dispersion_general}\\
\end{equation}
Similar to Ref.\cite{Fetter86} we will use the notation $\alpha_{0m}$ for mth zero of the function $J'_{0}(x)$. Using Eq.(\ref{L_C}) we re-write Eq.(\ref{Dispersion_general}) as it follows
\begin{equation}
\omega= v_{p}\frac{\alpha_{0m}}{R}\cdot \left [ \frac{1-e^{-2hq}}{2hq} \right ]^{1/2},
\label{Dispersion_detailed}\\
\end{equation}
where $v_{p}=\sqrt{\frac{4\pi e^{2} n h}{m^{*}\varepsilon}}$ is plasma wave velocity.

For strong screening case $qh \ll 1$ we obtain the relationship
\begin{equation}
\omega=v_{p}\frac{\alpha_{0m}}{R},
\label{Dispersion_gated}\\
\end{equation}
which is exactly the linear dispersion law\cite{Chaplik72,Fetter86} if one defines wave vector
\begin{equation}
q=\frac{\alpha_{0m}}{R}
\label{wave_vector}\\
\end{equation}
for present case of axisymmetric plasmon excitations.

In the opposite ungated case $qh \gg 1$ Eq.(\ref{Dispersion_detailed}) provides familiar long-wavelength dispersion law\cite{Stern67}
\begin{equation}
\omega=\sqrt{\frac{2\pi e^{2}n}{m^{*}\varepsilon}q}
\label{Dispersion_ungated}\\
\end{equation}
with the wave vector specified by Eq.(\ref{wave_vector}) as well.

Let's do visualization of the transition from ungated to screened 2D system by fixing the disk radius and, then varying the slab height.
It is convenient to introduce the screening parameter $\kappa=\sqrt{R/2h}$ and reference value of frequency
\begin{equation}
\omega_{0}(\varepsilon)=\sqrt{\frac{2\pi e^{2}n}{m^{*}\varepsilon R}}.
\label{Omega_reference}\\
\end{equation}

Using Eq.(\ref{Dispersion_detailed}) we plot in Fig.\ref{Fig1} the dimensionless frequency $\Omega=\omega/\omega_{0}(\varepsilon)$ vs $\kappa$ for lowest axisymmetric modes $m=2,3$. For highly screening case $\kappa \gg 1$ one obtains $\Omega=\frac{\alpha_{0m}}{\kappa}$ being a simple replica of  Eq.(\ref{Dispersion_gated}). For ungated 2D gas  $\kappa\ll 1$ we recover the result specified by Eq.(\ref{Dispersion_ungated}) as $\Omega=\sqrt{\alpha_{0m}}$. The change in axisymmetric mode behavior occurs at $\kappa\sim \sqrt{\alpha_{om}}$ which was not clearly indicated in longstanding classical study\cite{Fetter86}.

The second part of the present paper concerns more interesting case of partially gated 2D electron system shown in Fig.\ref{Fig1}, inset. Let the height of the plate $h$ be fixed. Obviously, the extension of the central gate spot would result in a transition from ungated to fully gated 2D gas. We now search the spectra of axisymmetric excitations by varying the central gate radius $0 \leq r_{0}\leq R$.

For gated disk(index "1") and ungated ring(index "2") the in-plane potential can be written by Eq.(\ref{Solution_Bessel}) taking into account the respective inductances and capacitances specified by Eq.(\ref{L_C}).
\begin{eqnarray}
U_{1}=A_{1}J_{0}(\rho_{1}) \qquad \qquad \qquad
\label{Potential_Gated}\\
U_{2}=A_{2}J_{0}(\rho_{2})+B_{2}Y_{0}(\rho_{2})
\label{Potential_Ungated}\\
U_{1}=U_{2} \mid _{r=r_{0}}, U'_{1}=U'_{2} \mid _{r=r_{0}}, U'_{2} \mid _{r=R}=0,
\label{Boundary_conditions}
\end{eqnarray}
where the dimensionless variables $\rho_{1,2}=r/\lambda_{1,2}$ are different for gated disk and ungated ring since $\lambda_{1,2}=\frac{1}{\omega \sqrt{LC_{1,2}}}$. The actual experiments data\cite{Kukushkin17,Kukushkin18,Kukushkin23} match the requirement $qh<<1$ valid for both the gated disk and the ungated ring. Using Eq.(3) we find $C_{1}=\varepsilon r/2h$ for the gated part. However, for ungated part one needs to account the permittivity of the air above the ring. For actual $qh<<1$ case the rigorous analysis\cite{Sydoruk19} provides $C_{2}=\tilde{\varepsilon}qr$, where $\tilde{\varepsilon}=(1+\varepsilon)/2$ denotes the effective dielectric permittivity.

Equations(\ref{Potential_Gated}),(\ref{Potential_Ungated}) have to be solved under certain boundary conditions. As before, we will assume a zero current at both the circumference and the center of 2D disk. The boundary conditions for interface between the gated and ungated parts are of special interest. Recall that the conventional LC-approach\cite{Aizin12} concerns a two-dimensional stripe along, for example, x-direction. Theory states\cite{Brews87} that both the in-plane voltage $U$ and the average complex power $P$ carried by plasmon wave are constants across the interface between gated-to-ungated parts. The power is, in turn, determined by the longitudinal component of the Poynting vector $S_{x}=c[\vec{E}\vec{H}]_{x}/4\pi$ averaged over the oscillation period. Here, $\vec{H}$ denotes the transverse magnetic field induced by plasmon propagating along two-dimensional stripe. Obviously, two-dimensional disk is not a case. Indeed, for arbitrary radial current the magnetic field is zero. Hence, the boundary conditions at the interface between the gated and ungated parts must be different compare to stripe case. It is natural and physically sensible to assume the potential and current as constants at the interface which is depicted by Eq.(\ref{Boundary_conditions}). Using the set of Eqs.(\ref{Potential_Gated},\ref{Potential_Ungated},\ref{Boundary_conditions}) we obtain the transcendental dispersion equation

\begin{eqnarray}
J_{0}(\Omega \Lambda \kappa \nu)(J_{1}(\Omega^{2} \nu)Y_{1}(\Omega^{2})-J_{1}(\Omega^{2})Y_{1}(\Omega^{2} \nu))+ \qquad
\label{Dispersion_partial_full} \\
\frac{k\Lambda}{\Omega}J_{1}(\Omega \Lambda \kappa \nu)\left [J_{1}(\Omega^{2})Y_{0}(\Omega^{2} \nu)-J_{0}(\Omega^{2} \nu)Y_{1}(\Omega^{2}) \right ]=0,
\nonumber
\end{eqnarray}
where $\Omega=\omega/\omega_{0}(\tilde{\varepsilon})$ is the dimensionless frequency, $\Lambda=\sqrt{\varepsilon/\tilde{\varepsilon}}$ is the auxiliary coefficient. Eq.(\ref{Dispersion_partial_full}) can be minimized as it follows
\begin{equation}
D_{U}\cdot \Re_{I}+\frac{k\Lambda}{\Omega}D_{I}\cdot \Re_{U}=0.
\label{Dispersion_partial}\\
\end{equation}
Eq.(\ref{Dispersion_partial}) allows one to obtain the plasmon frequency $\Omega$ vs dimensionless ratio $\nu=r_{0}/R$ at certain value of parameters $\kappa,\Lambda$. Here, we use the notations $D_{I}=J_{1}(\Omega {\bf \Lambda}\kappa \nu)$ and $D_{U}=J_{0}(\Omega {\bf\Lambda} \kappa \nu)$ for gated disk(D) part. The solutions $D_{I(U)}=0$ correspond to zero current at the disk center and, then the zero current(voltage) respectively at the disk circumference. The remaining multipliers, namely $\Re_{I}=J_{1}(\Omega^{2} \nu)Y_{1}(\Omega^{2})-J_{1}(\Omega^{2})Y_{1}(\Omega^{2} \nu)$ and $\Re_{U}=J_{1}(\Omega^{2})Y_{0}(\Omega^{2} \nu)-J_{0}(\Omega^{2} \nu)Y_{1}(\Omega^{2})$ correspond to ungated ring($\Re$) part seen in Fig.\ref{Fig1}, inset. The transcendental equations $\Re_{I(U)}=0$ define the solution of Bessel Eq.(\ref{Bessel}) for zero current(voltage) respectively at the inner boarder and, then the absence of the current at the outer circumference of the ring.

\begin{figure}[tbp]
\begin{center}\leavevmode
\includegraphics[width=1.0\linewidth]{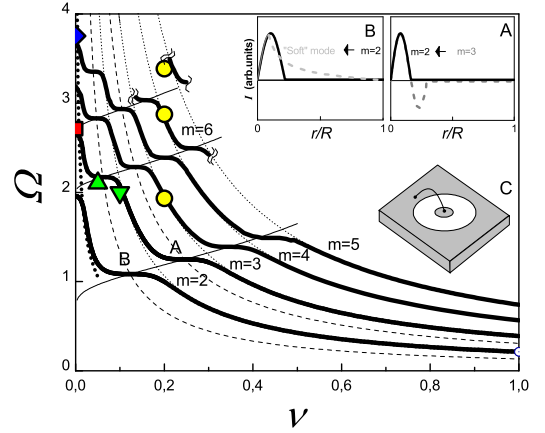} \caption[]{\label{Fig2} The dimensionless frequency $\Omega=\omega/\omega_{0}(\tilde{\varepsilon})$ vs
ratio $\nu=r_{0}/R$ for lowest $m=2..5$ and, partially $m=6,8$ plasma modes plotted for fixed $\kappa=18,\Lambda=1.36$. Arrows depict the frequency $\Omega=\sqrt{\alpha_{0m}}$, m=2..5 for totally ungated 2DEG. Dotted lines represent the asymptotes $\Omega=\frac{\alpha_{0m}}{\kappa \Lambda \nu}$ expected for plasmon localized in the gated part. Dashed(solid thin) lines depict $D_{U}=0$ and $\Re_{U}=0$ solutions respectively. Bold dotted line specifies the "soft" mode asymptote $\Omega=\frac{1}{\sqrt{\kappa \Lambda \nu}}$. The experimental data for ungated ($\square$\cite{Kukushkin18},$\lozenge$\cite{Kukushkin23}) and partially gated ($\bigtriangleup,\bigtriangledown,\bigcirc$\cite{Kukushkin17}) 2D systems is denoted in Table 1. Inset A(B): Radial distribution(schematically) of the current density for and intermode transition A and B respectively shown in the main panel. Inset C: Feedback 2D plasmon resonator(under Ref.\cite{Kukushkin20}).}
\end{center}
\end{figure}

For real systems\cite{Kukushkin18} the typical disk size $R=0.25$mm is much grater that the gate to 2D gas separation is $h=370$nm, hence we find $\kappa=18\gg 1$. Then, for GaAs/AlGaAs 2D system $\varepsilon=12.8$ one obtains $\Lambda=1.36$. Assuming that $\Omega \sim 1$ we conclude that second summand in Eq.(\ref{Dispersion_partial}) prevails. Therefore, the zeroth of transcendental equation $D_{I}=0$ basically defines the solution of Eq.(\ref{Dispersion_partial}). The later condition corresponds to axisymmetric plasmon localized in a gated part. The respective frequencies are given by asymptotes $\Omega=\frac{\alpha_{0m}}{\kappa {\bf \Lambda} \nu}$ shown by dotted lines in Fig.\ref{Fig2}. However, this simple scenario fails when the multipliers from first(second) summand in Eq.(\ref{Dispersion_partial}) vanish simultaneously $D_{U},\Re_{U}=0$ providing the dispersion Eq.(\ref{Dispersion_partial}) becomes fulfilled as well. The solutions of equations $D_{U}=0$ and $\Re_{U}=0$ are shown in Fig.\ref{Fig2} by thin and dashed lines respectively. The condition $D_{U},\Re_{U}=0$ exactly defines the transition between neighboring axisymmetric modes. To support our reasoning we plot schematically in Fig.\ref{Fig2}, inset A the spatial distribution of the current in a disk for initial $m=3$ and, then final $m=2$ neighboring plasma modes. Actually, the transition $3\rightarrow 2$ depicted as A in Fig.\ref{Fig2} can be viewed as a plasma wave that loses half of the period.

For small gate sizes when the most place atop the slab is occupied by ungated ring, the lowest mode $m=2$ undergo further transformation.
The point of interest B in Fig.\ref{Fig2} is still defined by condition $D_{U},\Re_{U}=0$. However, in the present case the
plasmon mode $m=2$ is first confined within ultra-narrow gated part and, then is modified entirely by occupation of the ungated part as whole. Let us
call further this mode as "soft" mode. In Fig.\ref{Fig2} we represent schematically the current spatial distribution for initial $m=2$ mode and,
finally, the "soft" mode. We verify the dispersion relation for "soft" mode follows the asymptote $\Omega=\frac{1}{\sqrt{\kappa \Lambda \nu}}$ shown by bold dotted line in Fig.\ref{Fig2}. In actual fact, the appearance of the "soft" mode is a precursor of the conventional plasmon localized in completely ungated 2D gas, i.e. when $\Omega=\sqrt{\alpha_{0m}}$.

It is worthwhile to mention with respect to Fig.\ref{Fig2} that upon change of the gate radius $r_{0}$ the plasmon frequency follows either $\omega\sim 1/r_{0}$ or $\omega\sim 1/\sqrt{R}$ dependencies regarding to the actual $\nu$-range of interest. Evidently, this uncertainty may lead in misunderstanding regarding accurate identification of the plasmon modes. We argue that the present model would be useful for analyzing the axisymmetric plasmon excitations persisting in partially gated 2D disk feedback system\cite{Kukushkin20} shown in Fig.\ref{Fig2},inset.

\begin{table}[h]
\begin{center}
\label{Table} \caption{Axisymmetric plasmon: theory vs experiment}
\begin{tabular} {@{}*{10}{l}}
 \hline\hline Samp. &$n\cdot 10^{11}$ & $h$ & $r_{0}$ & $R$ & $f_{0}$ & A & $f_{exp}$ & $\Omega_{exp}$ & Ref.\\
         &[cm$^{-2}$] & [nm] & [mm] & [mm] & [GHz] & & [GHz] & & \\
 \hline  N1 & 2.6 & -   & 0    & 0.25 & 30.0 & 0.31 & 60.0 & 1.0 & $\square$\cite{Kukushkin18}  \\
         N2 & 1.0 & 370 & 0.05 & 1.0  & 9.3  & 0.38 & 14.4 & 1.6 & $\bigtriangleup$\cite{Kukushkin17}\\
         N3 & 1.0 & 370 & 0.05 & 0.50 & 13.2 & 0.27 & 19.4 & 1.5 & $\bigtriangledown$\cite{Kukushkin17}\\
         N4 & 1.0 & 370 & 0.05 & 0.25 & 18.6 & 0.19 & 26.5 & 1.4 & $\bigcirc$\cite{Kukushkin17} \\
         N4 & 1.0 & 370 & 0.05 & 0.25 & 18.6 & 0.19 & 39.4 & 2.1 & $\bigcirc$\cite{Kukushkin17} \\
         N4 & 1.0 & 370 & 0.05 & 0.25 & 18.6 & 0.19 & 46.4 & 2.5 & $\bigcirc$\cite{Kukushkin17} \\
         N5 & 1.0 & -   & 0    & 0.50 & 13.2 & 0.27 & 36.4 & 2.8 & $\lozenge$\cite{Kukushkin23} \\
         N5 & 1.0 & -   & 0    & 0.50 & 13.2 & 0.27 & 47.0 & 3.6 & $\lozenge$\cite{Kukushkin23} \\
\hline\\
\end{tabular}
\end{center}
\end{table}

Let us compare our results with experimental data for AlGaAs/GaAs samples demonstrated\cite{Kukushkin17,Kukushkin18,Kukushkin23} axisymmetric plasmon excitations. At first, we calculate the reference frequency $f_{0}=\omega_{0}(\tilde{\varepsilon})/2\pi$ for each specimen presented
in Table 1. To confirm validity of our quasistatic approach, we calculate the ratio of the long-wavelength plasma frequency to light frequency with the same wave vector, $A=\omega_{0}(\tilde{\varepsilon})\sqrt{\varepsilon}R/c$, called retardation parameter\cite{Kukushkin03}. For all samples in Table 1 $A \ll 1$. We conclude that retardation effects are insignificant, therefore the use of non-retarded Poisson's equation is well justified.

Continuing the analysis of experimental data we calculate the dimensionless ratio $\Omega_{exp}=f_{exp}/f_{0}$ for each sample. The resulting data points are then added to the graph of the plasmon spectra in Fig.\ref{Fig2}. Remarkably, the experimental values match perfectly the certain modes of axisymmetric plasmon spectrum. For samples N1,2 the lowest observed plasmon mode corresponds to $m=2$, then the plasmon in sample N3 exhibits $m=3$ mode. The sample N4 demonstrated a series of the plasmon modes $m=4,5,6$. The ungated sample N5 exhibited the plasma mode for $m=3,5$. For samples N3-5, expected strong the fundamental axisymmetric mode $m=2$ was not observed. Although this finding looks puzzling, the answer is unpretentious. Recall that for dark axisymmetric plasmon excitation the coupling with external electromagnetic radiation is much weaker\cite{Kukushkin17} compared to conventional plasmons with nonzero angular momentum $l\neq 0$. Evidently, the dark modes would be hidden out by conventional ones. Indeed, the conventional $l=1$ modes observed\cite{Kukushkin18} at frequencies 18GHz and 13GHz cover the spectral range of axisymmetric fundamental modes expected at 14.2GHz  and 13,7GHz for samples N4 and N3 respectively. Finally, for sample N5 one expects the fundamental dark mode at 25.6GHz which close to\cite{Kukushkin23} with conventional mode $l=1$ at 19GHz.

In conclusion, we investigated the transition from ungated to gated two-dimensional system of disk geometry by means of central spot gate expansion. The discontinuity of gated to ungated part provides the interchange between the neighboring axisymmetric modes. The experimental data for different sample sizes, carrier densities is found to agree well with theory predictions. Our study allows to classify the observed dark axisymmetric modes and, moreover, pave a
way to feedback plasmon resonator\cite{Kukushkin20} study.

\bibliography{AXIS_PLASMON}

\end{document}